\documentstyle[prl,aps,preprint]{revtex}
\begin{document}

\title{ Quantum melting of the long-range antiferromagnetic
order and spin-wave condensation in $t-J-V$ model }

\author {M.Yu. Kuchiev$^{a}$ and O. P. Sushkov$^{b}$}
\address { School of Physics, The University of New South Wales
 Sydney 2052, Australia}

\date{June 26, 1994}
\maketitle

\begin{abstract}
We consider two-dimensional $t-J-V$ model. The wave function of the ground
state is constructed. We demonstrate that the doping by holes results in
condensation of the spin-waves, destruction of the long-range antiferromagnetic
order and formation of the gap in the spin-wave spectrum.
\end{abstract}

\vspace{0.5cm}
\hspace{3.cm}PACS numbers: 71.27.+a, 74.20.Hi, 75.50.Ee
\vspace{1.cm}

It is well known that the long-range antiferromagnetic (AF) order in Cooper
Oxide superconductors is destroyed under the doping by the holes.
In the frameworks of two-dimensional
$t-J$ model the origin of this instability was realized.
(see e.g. Refs.\cite{Shr9,Dom0,Sin0,Ede1,Iga2,Sus3}). The instability
is due to the strong interaction of spin-waves with mobile holes. However
a structure of the ground state as well as a spectrum of excitations was
not understood. In the present work we discuss the ways to stabilize the
hole-hole pairing, construct the ground state wave function as a condensate
of spin-waves and discuss the spectrum of bosonic excitations.

At half-filling (one hole per site) the $t$-$J$ model is equivalent to
the Heisenberg AF model which has long-range
AF order in the ground state.
We consider the doped system starting from this ground state
which we denote by $|0\rangle$. In spite of the destruction of
the long-range AF order it is convenient to use $|0\rangle$
and corresponding quasiparticle excitations as a basis set in the problem
with doping.

The effective Hamiltonian for
$t-J$ model was derived in works\cite{Suhf,Cher3,Kuch3}
\begin{equation}
\label{1}
H_{eff}=\sum_{{\bf k}\sigma}\epsilon_{\bf k}h_{{\bf k}\sigma}^{\dag}
h_{{\bf k}\sigma}+
\sum_{\bf q}\omega_{\bf q}(\alpha_{\bf q}^{\dag}\alpha_{\bf q}
+\beta_{\bf q}^{\dag}\beta_{\bf q}) +H_{h,sw} + H_{hh}.
\end{equation}
It is expressed in terms of usual spin-waves on AF
background $\alpha_{{\bf q}}, \beta_{{\bf q}}$
(see e.g. Ref.\cite{Manousakis}), and
composite hole operators $h_{{\bf k}\sigma}$ ($\sigma = \pm 1/2$).
The summations over ${\bf k}$ and ${\bf q}$ are restricted inside the Brillouin
zone of one sublattice where
$\gamma_{\bf q}= {1\over 2} (\cos q_x+\cos q_y)\ge 0$.
The spin-wave dispersion is $\omega_{\bf q}=2\sqrt{1-\gamma_{\bf q}^2}
\approx \sqrt{2}|{\bf q}|, \ at \ q \ll 1$. Let us recall that the
parameters of $t-J$ model are $t$ and $J$. We set hereafter $J=1$, so
all energies are measured in units of $J$.
Single hole dispersion has minima at the face of magnetic Brillouin zone
${\bf k}_0=(\pm \pi/2, \pm \pi/2)$. Near these points
the dispersion can be presented in a usual quadratic form
$\epsilon_{\bf p}\approx{1\over2}\beta_1 p_1^2+
{1\over2}\beta_2 p_2^2$ ($\beta_2 \ll \beta_1$),
where ${\bf p}={\bf k}-{\bf k}_0$, and
$p_2$ is projection along the face of Brillouin
zone, $p_1$ is orthogonal projection of the momentum.
For $5\ge t \ge 1$ $\beta_1 \approx 0.65t$ (see e.g.
Refs.\cite{Mart1,Sus2,Giam3}).
Following Refs.\cite{Mart1,Giam3} we will set  $a=\beta_1/\beta_2
\approx 5-7$.
For the small concentrations $\delta \ll 1$ under consideration,
 holes are localized in momentum space in the vicinity of the minima of the
 band  and  the Fermi surface consists of ellipses.
The Fermi energy and Fermi momentum of non-interacting holes are:
$ \epsilon_F= \frac{1}{2} \pi (\beta_1 \beta_2)^{1/2} \delta$,
$p_F \sim (\pi \delta)^{1/2}$.
The Fermi momentum $p_F$ is measured
  from the center of the corresponding ellipse.
Let us stress that the numerical value of $\epsilon_F$ is very small.
For realistic superconductors $t/J \approx 3$
 (see, e.g., Refs.\cite{Esk0,Fla1}).
Therefore at $\delta=0.1$ and $J=0.15$ eV one gets
 $\epsilon_F\approx 15$ meV $\approx 175$ K.

 The effective interaction of a composite hole with a
spin-wave is of the form (see, e.g. Refs.\cite{Mart1,Suhf})
\begin{equation}
\label{2}
H_{h,sw}= \sum_{{\bf k},{\bf q}}g({\bf k},{\bf q})
\biggl(h_{{\bf k}+{\bf q}\downarrow}^{\dag}
h_{{\bf k}\uparrow} \alpha_{\bf q}+h_{{\bf k}+{\bf q}\uparrow}^{\dag}
h_{{\bf k}\downarrow} \beta_{\bf q} + H.c. \biggr).
\end{equation}
For ${\bf k} \approx {\bf k}_0$ and $q \ll 1$ the vertex is
$g({\bf k},{\bf q})\approx 2^{3/4}fq_1/\sqrt{q}$.
The component $q_1$ is perpendicular to the face of Brillouin zone.
The coupling constant $f$
was calculated in the Ref.\cite{Suhf}. For $t=3$ it is equal
$f\approx 1.8$, for large $t$ the coupling constant $f$ approaches to 2.

  The interaction between the two holes can be caused by the exchange
of single  spin-wave. Alongside with that there is a
contact hole-hole interaction which is denoted in (\ref{1}) by
$H_{hh}$. It is of the form
\begin{equation}
\label{3}
H_{hh}\approx 8 \sum_{1,2,3,4}\biggl[ A\gamma_{{\bf k_1}-{\bf k_3}}
 + {C\over 2} (\gamma_{{\bf k}_1+{\bf k}_3}+\gamma_{{\bf k}_2+{\bf k}_4})
    \biggr]h_{3\uparrow}^{\dag}h_{4\downarrow}^{\dag}
h_{2\downarrow}h_{1\uparrow} \delta_{12,34}.
\end{equation}
An expressions for the coefficients $A$ and $C$ as a functions of $t$
are presented in the works\cite{Cher3,Kuch3}.

  It was demonstrated in the works\cite{Fla4,BCDS} that the spin-wave
exchange results in very strong pairing between the holes. The pairing
is strongest in d- and g-waves where the superconducting gap $\Delta$
is of the order of Fermi energy $\epsilon_F$. The approach
works\cite{Fla4,BCDS} is based on the observation that at the typical momentum
$q \sim p_F$ the spin-wave frequency is much larger then Fermi energy:
$\omega_q \sim p_F \sim (\pi \delta)^{1/2} \gg \epsilon_F =
{1\over{2}}(\beta_1\beta_2)^{1/2} \delta$.
This is why we can calculate pairing using unrenormalized spin-waves.
Now we are going to discuss what happens with spin-waves at $q < p_F$.
Let us consider renormalized spin-wave Green function.
\begin{equation}
\label{GF}
G(\omega,{\bf q})
={{1}\over{\omega^2-\omega_q^2+P(\omega,{\bf q})}}.
\end{equation}
For stability of the system the condition
\begin{equation}
\label{stab}
\omega_q^2 > P(0,{\bf q})
\end{equation}
should be fulfilled. Otherwise the Green function (\ref{GF}) would
possesses a pole with imaginary $\omega$.
The diagrams for polarization operator are
presented in Fig.1. Chaining is due to the contact interaction
(\ref{3}). In $t-J$ model the constant $C$ in contact interaction
is small at any $t$, and the constant $A$ vanishes exactly near the
point which we are interested in: $t \approx 3$. Therefore the
contact interaction is small and polarization operator is given
by first diagram only. If we assume the holes be a normal Fermi
liquid the calculation $P(\omega,{\bf q})$ is very simple. For
$q \ll p_F$ one gets\cite{Sus3}
\begin{equation}
\label{P0q}
P(0,{\bf q})\approx {{4f^2}\over{\pi\sqrt{\beta_1 \beta_2}}}q^2.
\end{equation}
 After the substitution of the values
of $f$ and $\beta_1, \beta_2$ presented above, we see that condition
(\ref{stab}) is violated, and $P$ is about 2.5 times larger than $\omega_q^2$.
This indicates the instability of normal Fermi liquid ground
state\cite{Shr9,Dom0,Sin0,Ede1,Iga2,Sus3}.
What happens if we take into account the hole-hole pairing?
In this case there is no simple analytical expression for the
polarization operator. Our numerical computations show that
pairing practically does not influence $P(0,{\bf q})$.
With pairing calculated in the works\cite{Fla4,BCDS} $P(0,{\bf q})$
is only by 8\% smaller then the value given by (\ref{P0q}).
Let us note that it is quite strong pairing: the maximal value of
superconducting gap on the Fermi surface is $\Delta \sim 0.7\epsilon_F$,
and this gives reasonable values of critical temperature.
Even if we enhance pairing by hands up to the value
$\Delta \sim 1.3\epsilon_F$, $P(0,{\bf q})$
is only by 14\% smaller then the value given by (\ref{P0q}).
Thus even with pairing we have instability of ground state. Let us
stress that this conclusion is practically independent of the
mechanism of pairing.

 One can believe that nonlinear
interaction of spin-waves which is not included into effective Hamiltonian
(\ref{1}) could stabilize the system. However the maximal magnitude of
imaginary $\omega$ is at $q \sim p_F$. Here the spin-wave residue of unstable
mode in Green function (\ref{GF}) is very small. It means that actually the
system is unstable with respect to spin-sound in hole Fermi liquid.
The picture is as follows: The Green function (\ref{GF}) has two
collective poles: 1)the upper pole which originates from initial spin-wave,
2)the lower pole which originates from spin-sound.
Due to the interaction the spin sound is repulsed down from initial
spin-wave. The repulsion is so strong that it acquires imaginary
frequency. The nonlinear interaction of spin-waves practically does
not influence the spin sound and therefore can not eliminate
the instability. We do not see any possibility
to stabilize $t-J$ model without contact interaction. The true
ground state in this case is probably some spiral phase.

 Let us introduce now the hole-hole repulsion $V$ at nearest sites and consider
$t-J-V$ model. In the effective Hamiltonian (\ref{1}) only contact
interaction (\ref{3}) is changed: we should consider the constant $A$
as independent parameter of the model ($A\approx V$). Now the chain in Fig.1
becomes essential, and simple estimations show that value
$A \sim 1.0 - 1.3$ is enough to eliminate
the instability. If $J=0.15 eV$ it means that $A \sim 0.2eV$.
It is hardly believed that in realistic systems there is no such
a small Coulomb repulsion between the holes at nearest sites.
The repulsion is probably even larger, but in this case, for the chain Fig.1
one has to use hole-hole scattering amplitude instead of
simple matrix element $H_{hh}$. Calculation of this amplitude is in
the progress.

 Consider now the problem of long-range AF order in $t-J-V$
model with $V$ big enough, so that the paired hole Fermi liquid is stable.
For this question it is convenient to use Hamiltonian technique instead of
Feynman one because we need explicit construction of ground state wave
function.
The polarization operator $P(\omega,{\bf q})$ in (\ref{GF}) corresponds to
normalization $2\omega_q$ spin-waves in the volume. For Hamiltonian
approach let us introduce the polarization operator
$\Pi(\omega,{\bf q})$ corresponding to normalization one spin-wave
in the volume: $P(\omega,{\bf q})=2\omega_q\Pi(\omega,{\bf q})$.
The wave function of renormalized spin-wave
corresponding to Green function (\ref{GF}) is a combination of
$\alpha_{\bf q}^{\dag}$ and $\beta_{\bf -q}$. To find this wave function
write down the effective spin-wave Hamiltonian.
\begin{equation}
\label{sw}
H_{sw}=\sum_{\bf q}\biggl((\omega_{\bf q}-\Pi(\omega,{\bf q}))
(\alpha_{\bf q}^{\dag}\alpha_{\bf q}+\beta_{\bf q}^{\dag}\beta_{\bf q})
+\Pi(\omega,{\bf q})
(\alpha_{\bf q}\beta_{\bf -q}+\alpha_{\bf q}^{\dag}\beta_{\bf -q}^{\dag})
\biggr).
\end{equation}
The term proportional to $\omega_{\bf q}$ comes from ``bare'' Hamiltonian
(\ref{1}).
First ``$\Pi$ term'' comes from diagram Fig.2a where one
spin-wave is annihilated and the other is created. (For simplicity we
do not present the chain with hole-hole contact rescattering)
Second ``$\Pi$ term'' comes from diagrams Fig.2bc where
two spin-waves are annihilated or created. Let us note that spin-waves
have definite values of $S_z$: $\alpha_{\bf q}^{\dag}$ has $S_z=-1$
and $\beta_{\bf -q}$ has $S_z=+1$. Therefore they can appear only in
combinations presented in (\ref{sw}). One can certainly prove this
explicitly using the vertex (\ref{2}) and calculating the polarization
operator. In the second ``$\Pi$ term'' the spin-waves have the opposite
momenta.
The vertex (\ref{2}) is proportional to the momentum. Just due to this
reason the second ``$\Pi$ term'' has different sign in
comparison with first one.
Dioganalization of Hamiltonian (\ref{sw}) by Bogoliubov transformation
gives the spectrum of Bose excitations in the system
\begin{equation}
\label{om}
\Omega_{\bf q}^2=(\omega_{\bf q}-\Pi)^2-\Pi^2=
\omega_{\bf q}^2-2\omega_{\bf q}\Pi(\Omega_{\bf q},{\bf q}).
\end{equation}
This is exactly the equation for the poles of Green function (\ref{GF}).
To find new ground state we have dioganalize (\ref{sw}) at $\omega=0$.
As usually for Bogoliubov transformation this ground state is of the form
\begin{equation}
\label{gs}
|gs\rangle \propto \exp \biggl(\sum_{\bf q}c_{\bf q}
\alpha_{\bf q}^{\dag}\beta_{\bf -q}^{\dag}\biggr)|0 \rangle.
\end{equation}
This is exactly the condensate of spin-waves.

  We started from Neel ground state $|0\rangle$ with two sublattices
$A$-up and $B$-down. The difference in magnetization of two
sublattices is of the form (see e.g. Ref.\cite{Taka})
\begin{equation}
\label{sz}
{1\over 2}(S_A^z-S_B^z)=1-f_0-2\sum_{\bf q}{1\over{\omega_{\bf q}}}
\biggl(\alpha_{\bf q}^{\dag}\alpha_{\bf q}+\beta_{\bf q}^{\dag}\beta_{\bf q}
-\gamma_{\bf q}
(\alpha_{\bf q}\beta_{\bf -q}+\alpha_{\bf q}^{\dag}\beta_{\bf -q}^{\dag})
\biggr),
\end{equation}
where $1-f_0 \approx 0.303$. Using parameters of transformation
dioganalizing (\ref{sw}) one can easily calculate
renormalized magnetization
\begin{equation}
\label{magn}
\langle gs|{1\over 2}(S_A^z-S_B^z)|gs\rangle=1-f_0-2\int
\biggl({1\over{\Omega_{0{\bf q}}}}-{1\over{\omega_{\bf q}}}\biggr)
{{d^2{\bf q}}\over{(2\pi)^2}},
\end{equation}
where $\Omega_{0{\bf q}}=
\sqrt{\omega_{\bf q}^2-P(0,{\bf q})}$
At small $q$ due to the superconducting gap
$\Omega_{0{\bf q}} \approx \Omega_{\bf q}$. At $q \gg p_F$ the
polarization operator vanishes and $\Omega_{0{\bf q}} \to \omega_{\bf q}$.
Therefore the integral in Eq.(\ref{magn}) converges at $q \sim p_F$,
and we get the estimation for $\delta S_z$ coursed by spin-wave
condensation
\begin{equation}
\label{ds}
\delta S_z \sim -\sqrt{{\delta}\over{2\pi}}{{v}\over{\tilde{v}}}.
\end{equation}
Here $v=\sqrt{2}$ is unrenormalized spin-wave velocity and $\tilde{v}$
is renormalize that which follows from Eq.(\ref{om}). If $|\delta S_z|
= 0.303$ the magnetization vanishes and one should conclude that the long
range AF order is destroyed. Due to estimation (\ref{ds})
for $v/\tilde{v} \approx 4$ it happens at $\delta =\delta_c \sim 0.04$.
Note that
the considered effect of enhancement of spin quantum fluctuations due
to the polarization of fermionic subsystem is similar to the well known
Casimir effect in Quantum Electrodynamics.

What happens if $\delta > \delta_c$  and magnetization calculated
using formula (\ref{magn}) becomes negative? It means that there are a
lot of spin-waves in condensate and we have to take into account their
nonlinear interaction. We can not do it exactly.
Fortunately there is a simple approximate way suggested by
Takahashi in the work on Heisenberg model at nonzero temperature\cite{Taka}.
Following Takahashi let us impose the condition that sublattice magnetization
vanishes.
\begin{equation}
\label{s0}
\langle gs|{1\over 2}(S_A^z-S_B^z)|gs\rangle=0.
\end{equation}
To find the ground state with this condition we have to dioganalize
\begin{equation}
\label{Hnu}
H_{\nu}=H_{sw}-{1\over8}\nu^2(S_A^z-S_B^z),
\end{equation}
where $H_{sw}$ is given by (\ref{sw}) and $(S_A^z-S_B^z)$ by (\ref{sz}).
Simple calculation shows that instead of (\ref{om}) we get a spectrum
of excitations with a gap
\begin{equation}
\label{gap}
\Omega_{\bf q}^{\nu}=\sqrt{\Omega_{\bf q}^2+\nu^2}.
\end{equation}
The average value of magnetization is given by the formula (\ref{magn})
with $\Omega_{0{\bf q}}^{\nu}=\sqrt{\Omega_{0{\bf q}}^2+\nu^2}$ instead of
$\Omega_{0{\bf q}}$. We have to
find the gap $\nu$ substituting this formula into condition (\ref{s0}).
Let us stress that this condition reflects strong
nonlinearity of theory. In essence it is effective cutoff of
unphysical states in Dyson-Maleev approach (see discussion in the
work\cite{Dot}). Certainly the suggested
solution is not exact. It is kind of variational approach

 From the Eqs.(\ref{magn}),(\ref{s0}),(\ref{gap}) we conclude
\begin{equation}
\nu \propto (\sqrt{\delta}-\sqrt{\delta_c}).
\end{equation}
However for detailed calculations of the spin-wave gap we have to take
into account
not only the ``Casimir'' contribution (\ref{magn}),(\ref{ds}) into spin
quantum fluctuation, but also the contribution which is due to the
spin-wave exchange in hole-hole pairing. Such detailed calculation
will be presented elsewhere.

  In this work we have introduced  the
short range hole-hole repulsion $V$. One can prove that very small
value of $V$ practically destroys the d-wave hole-hole pairing.
However it does not influence the g-wave pairing which has the
same long-range behaviour as d-wave, but quite different short-range
one\cite{Fla4,BCDS}. Therefore the presented scenario favours the
g-wave pairing.

{\bf ACKNOWLEDGMENTS}

  We are very grateful to V. V. Flambaum and A. V. Dotsenko for valuable
discussions. This work was supported in part by a grant of the
Australian Research Council.

\newpage
{\bf FIGURE CAPTIONS}

FIG. 1. Spin-wave polarization operator.\\

Fig. 2. Spin-wave polarization operator in Schrodinger
representation:\\
a)One spin-wave is annihilated and the other is created.\\
b,c)Two spin-waves are annihilated or created.

\begin{references}
\bibitem[a]{byline1}Also at the
A.F.Ioffe Physical-Technical Institute, 194021 St. Petersburg, Russia
\bibitem[b]{byline1} Also at the Budker Institute of Nuclear Physics,
630090 Novosibirsk, Russia
\bibitem{Shr9} B.Shraiman and E.Siggia, Phys. Rev. Lett. {\bf 62} (1989) 1564.
\bibitem{Dom0} T.Dombre, J.Physique  {\bf 51} (1990) 847.
\bibitem{Sin0} A.Singh and Z.Tesanovic, Phys. Rev. B {\bf 41} (1990) 614.
\bibitem{Ede1} R.Eder, Phys. Rev. B {\bf 43} (1991) 10706.
\bibitem{Iga2} J.Igarashi and P.Fulde, Phys. Rev. B {\bf 45} (1992) 10419.
\bibitem{Sus3} O.P.Sushkov and V.V.Flambaum, Physica C {\bf 206} (1993) 269.
\bibitem{Suhf} O. P. Sushkov, Phys. Rev. B {\bf 49} (1994) 1250.
\bibitem{Cher3} A.L.Chernyshev, A.V.Dotsenko, and O.P.Sushkov,
Phys. Rev. B {\bf 49} (1994) 6197.
\bibitem{Kuch3} M.Yu.Kuchiev and O.P.Sushkov, Physica C {\bf 218} (1993) 197.
\bibitem{Manousakis} E. Manousakis, Rev. Mod. Phys. {\bf 63} (1991) 1.
\bibitem{Mart1} G.Martinez and P.Horsch, Phys. Rev. B {\bf 44} (1991) 317.
\bibitem{Sus2} O.P.Sushkov, Solid State Communications {\bf 83} (1992) 303.
\bibitem{Giam3} T.Giamarchi and C.Lhuillier, Phys.Rev. B {\bf 47} (1993)
2775 .
\bibitem{Esk0} H. Eskes, G. A. Sawatzky, and L. F. Feiner, Physica C
{\bf 160} (1989) 424.
\bibitem{Fla1} V. V. Flambaum and O. P. Sushkov,
  Physica C {\bf 175} (1991) 347.
\bibitem{Fla4} V.V.Flambaum, M.Yu.Kuchiev and O.P.Sushkov, Physica C, to be
published.
\bibitem{BCDS} V.I. Belinicher, A.L.Chernyshev, A.V.Dotsenko, and O.P.Sushkov,
submitted to Phys. Rev. B.
\bibitem{Taka} M.Takahashi Phys. Rev. B {\bf 40} (1989) 2494.
\bibitem{Dot} A.V.Dotsenko and O.P.Sushkov, to be published.



\end{references}
\end{document}